\documentclass[prd,showpacs,superscriptaddress,twocolumn,floatfix,preprintnumbers,altaffilletter, letterpaper]{revtex4-1}

\usepackage{xcolor}
\usepackage{graphicx}  
\usepackage{multirow}

\usepackage{enumitem} 
\graphicspath{{.} {figures}}
\graphicspath{{figures/}}
\usepackage{lipsum}
\usepackage{notes2bib}
\usepackage[colorlinks=true,citecolor=teal, linkcolor=blue]{hyperref}
\usepackage{url}

\usepackage{booktabs}

\usepackage[T1]{fontenc}
\usepackage{subcaption}
\usepackage{comment} 

\usepackage{amsmath,amssymb}   

\usepackage{amsfonts} 

\usepackage{amsmath,booktabs}

\begin{document}

\title{First demonstration of neural sensing and control in a kilometer-scale gravitational wave observatory}

\author{
N. Mukund,$^{1,2,3}$
J. Lough,$^{1}$ 
A. Bisht,$^{1}$ 
H. Wittel ,$^{1}$
S. Nadji ,$^{1}$
C. Affeldt ,$^{1}$
F. Bergamin,$^{1}$ 
M. Brinkmann ,$^{1}$
V. Kringel ,$^{1}$
H. L{\"u}ck ,$^{1}$
M. Weinert ,$^{1}$
K. Danzmann ,$^{1}$\\
$^{1}$Max-Planck-Institut f{\"u}r Gravitationsphysik (Albert-Einstein-Institut) and Institut f{\"u}r Gravitationsphysik, \\ Leibniz Universit{\"a}t Hannover, Callinstra{\ss}e 38, 30167 Hannover, Germany \\
$^{2}$LIGO, Massachusetts Institute of Technology, Cambridge, MA 02139, USA\\
$^{3}$NSF AI Institute of Artificial Intelligence and Fundamental Interactions, \\Massachusetts Institute of Technology, Cambridge, MA 02139, USA
}


\date{\today}

\begin{abstract}

Suspended optics in gravitational wave (GW) observatories are susceptible to alignment perturbations, particularly slow drifts over time, due to variations in temperature and seismic levels. Such misalignments affect the coupling of the incident laser beam into the optical cavities, degrade both circulating power and optomechanical photon squeezing and thus decrease the astrophysical sensitivity to merging binaries. Traditional alignment techniques involve differential wavefront sensing using multiple quadrant photodiodes but are often restricted in bandwidth and are limited by the sensing noise. We present the first-ever successful implementation of neural network-based sensing and control at a gravitational wave observatory and demonstrate
low-frequency control of the signal recycling mirror at the GEO\,600 detector. Alignment information for three critical optics is simultaneously extracted from the interferometric dark port camera images via a CNN-LSTM network architecture and is then used for MIMO control using soft actor-critic-based deep reinforcement learning. Overall sensitivity improvement achieved using our scheme demonstrates deep learning's capabilities as a viable tool for real-time sensing and control for current and next-generation GW interferometers.

\end{abstract}
\pacs{}
\maketitle

\section*{} \label{Intro}

\noindent GEO\,600 is a dual recycled advanced Michelson interferometer (IFO) with folded arms \cite{Grote_2004,H_Lueck_2010,Dooley_2016}, located near Hannover, Germany. With a peak strain sensitivity of about $10^{-22}/\sqrt{\text{Hz}}$ at 1 kHz, the observatory operates in AstroWatch mode \cite{Grote_2010} and takes astrophysically relevant gravitational wave (GW) data in the frequency band of 40\,Hz to 6\,kHz. In April 2020, GEO completed a joint observation with the KAGRA detector \cite{10.1093/ptep/ptab018,Somiya_2012} and searched for transient GW events from neutron-star binaries and generic unmodeled transients \cite{ligo2022first}. Several technologies pioneered at GEO\,600 \cite{Affeldt_2014} have been adopted at Advanced LIGO \cite{Aasi_2015,PhysRevD.93.112004} and Advanced Virgo \cite{Acernese_2015} and have played a crucial role in advancing GW instrumentation science. One such example is the continuous application of non-classical sources of light \cite{PhysRevLett.110.181101}, and the demonstration six \,dB of measured optical squeezing \cite{PhysRevLett.126.041102} on the key future upgrade goals for gravitational wave detectors.

In this work, we present another novel technique using neural networks (NNs) and demonstrate their capabilities to sense and control the state of the interferometer. The paper is organized as follows: Sec. \ref{AA} describes the existing alignment scheme and its limitations, Sec. \ref{NeuralSensing} explains the motivations and architecture of the neural sensing, Sec. \ref{Controller} describes the implementation of a deep reinforcement learning (RL) based controller and  Sec. \ref{RESULTS} provides the network predictions and improvements to the sensitivity when the trained controller is deployed for low-frequency signal recycling alignment control.

\section{Automatic Alignment} \label{AA}

\noindent  The astrophysical sensitivity of a Michelson interferometer can be improved by including two extra cavities, a power recycling cavity (PRC) and a signal recycling cavity (SRC), leading to an improved signal-to-noise ratio in the readout channel \cite{meers1988recycling}. 
The PRC at GEO\,600 consists of the PR mirror, located at the IFO's input port, and the Michelson IFO. With an optical gain of about 800, it is used to enhance the circulating laser power leading to a reduced level of the photon shot noise. Similarly, the SRC is formed by the SR mirror, situated at the IFO's output port, and the Michelson IFO (shown in Figure \ref{fig:GEO_layout}). It complements the PRC by forming a resonant cavity to enhance the signal sidebands from potential GWs. The SR mirror's microscopic position also determines the cavity's overall frequency response. The light that leaks out in transmission of the SR mirror is filtered using the output mode cleaner (OMC) and is sent to the main photodiode, which is then calibrated to produce the final GW strain data.

The IFO mirrors are suspended as multistage pendulum assemblies to suppress the seismic noise coupling, with the PR mirror having two pendulum stages and the Michelson mirrors and the SR mirror having three. However, the noise suppression is achieved only above the pendulum's resonance frequency, which is close to 1\,Hz. While this isolation is sufficient within the gravitational wave measurement band, the residual pendulum motion around the resonance frequency can cause misalignment of mirrors and long-term drifts, which is detrimental to the required sensitivity of the interferometer. Sub-optimal alignment of the incident beam to the OMC leads to intensity fluctuations in the photodiode signal, degrading the overall optical gain and increasing glitches that often mimic the actual GW signal. Such misalignments also routinely interfere with the suite of optical squeezing and thermal compensation experiments carried out at GEO. Automatic alignment systems are hence critical to attain optimal sensitivity and maintain long lock stretches at the observatory.

\par The goal of the auto-alignment system is to keep the axis of an incoming beam aligned to that of the cavity axis. In addition, it also keeps the beam spots centered on the mirrors. Angular alignment of the IFO mirrors is primarily carried out using the differential wavefront sensing technique (DWS) \cite{Morrison:94, Morrison:94_2} and becomes active once the cavities are \lq locked\rq\,\, in length using the PDH technique \cite{PDH}. In the DWS technique, phase modulation is imprinted onto the beam incident on a cavity, which is promptly reflected. It is then superimposed over another light field that leaks out of the cavity. This combined light field falls on a pair of quadrant photodetectors that are placed with a relative Gouy phase of 90$^{\circ}$. The angle and displacement between the two beams are obtained by taking the difference of photocurrent (demodulated at the modulation frequency) from the different QPD sections. If the beam spot is off-center by one beam radius, then about $86 \% \;(1-e^-{2})$ of the DWS signal is lost \cite{Grote:2003ypa}. Hence, there are usually two additional auxiliary centering control loops for DWS, one associated with each quadrant photodetector, that keep the beam spots centered on it. We use additional spot position control loops to keep these beam spots centered on each mirror. In the transmission port of each mirror is a quadrant photodetector that looks at the position of the beam spot. The cavity mirrors are then actuated directly or in some combination of available external actuators (preceding suspensions) to achieve the desired spot position control.

\par The DWS control has a bandwidth of up to 6 Hz, while the centering control loops are the fastest, having up to 1 kHz bandwidth. The slowest is the spot position control loops having less than 0.1 Hz bandwidth. For completeness, we would also like to mention that waist-position and waist-size mismatch between interfering beams are second-order misalignments that are not actively controlled but by optimal layout design. Despite the auto-alignment system, residual mirror misalignments can couple directly to the strain signal or through the interlinked cavities. One well-known mechanism is bilinear noise coupling \cite{PhysRevD.101.102006}, where the Michelson misalignment couples via the SR longitudinal degree. Such a coupling pathway exists since the PRC and SRC share the Michelson. Error signals for the DWS generated via Schnupp modulation result in the creation of radio frequency (RF) sidebands which are tens of MHz offset with respect to the laser(or primary carrier) frequency. Although the OMC suppresses these MHz sidebands and the higher-order-modes by a factor of 100 beyond its optical bandwidth at 2.9 MHz, they still leak into the final photodiode signal leading to an elevated shot noise floor and a reduced level of optical squeezing. Decreasing the level of RF sidebands is not viable with the existing system as it leads to a low SNR error signal, making it harder to control. Additionally, environmental events like excessive seismic motion or thermal fluctuations introduce sensing noise leading to off-centering or clipping of the beam on the DWS photodiode, impacting the drift control loops, often requiring a manual inspection. Consequently, the DC position of all the mirrors, notably the SR mirror, has to be tuned once a week for optimal detector sensitivity.

Another alternative to DWS in use at GEO is the dithering scheme. It involves mechanically oscillating the relevant optics at a specific frequency for each degree of freedom. The transmitted cavity power recorded by a single-element photodiode is then demodulated at the respective frequency to infer the corresponding misalignment. Such a scheme is used to align, for example, the OMC by dithering one of the beam-directing optics. This scheme has a lower bandwidth (20\,mHz) and causes additional jitter on the incident beam, leading to a 0.2 dB loss of squeezing. An alternative scheme based on modulated differential wavefront sensing is currently under commissioning for the OMC alignment \cite{bisht2020modulated}. The dithering also enhances the bilinear coupling to the strain if the beam is not well centered on the optic. All these reasons motivate the need for a better solution.

\section{Neural Sensing} \label{NeuralSensing}

\subsection{Why darkport is a good witness}
\noindent The south port of the IFO, referred to as the darkport, is usually kept close to destructive interference but with a slight DC offset of about 5-50 picometers. This offset allows about six mW of carrier to leak out and about 30 mW of higher-order modes to exit via the darkport. The higher-order modes originate inside the IFO due to mismatch in the interfering beams, which are caused by thermal lensing of the beam splitter,  microscopic imperfections on the mirror surfaces, or residual misalignment of the mirrors. Consequently, video camera images of the darkport (DP) beam contain much information about the IFO state. It is used for manual pre-alignment, making the longitudinal lock acquisition easier, and then the wavefront sensor-based auto-alignment systems take over. In the lock, the DP image shows breathing motion corresponding to the residual movement of the suspended optics. A skilled commissioner can often judge some alignment states from this image.

\par The error signals of several feedback loops can broadly determine the state of the IFO. In particular, the Michelson differential, PRC, and the SRC alignment DOFs play a crucial role for GEO. Sensing noise entering the existing DWS-based scheme is often not sufficiently corrected by the current control loops, leading to pointing drifts and sensitivity degradation. Timescales of these disturbances range from hundreds of milliseconds to a few days and include sources like temperature variations, seismic disturbances, and optomechanical intra-cavity cross-couplings. However, since a strong mapping exists between the interferometer's state and the darkport image, such disturbances are encoded in their breathing patterns. Once a week, these loop offsets are tuned by commissioners via visual inspection of the camera images. The values are considered optimized when the darkport image resembles a stable state, often based on recollections from memory.

\subsection{Coherence Mapping}

\begin{figure}[!htb]
    \centering
    \includegraphics[width=0.9\linewidth]{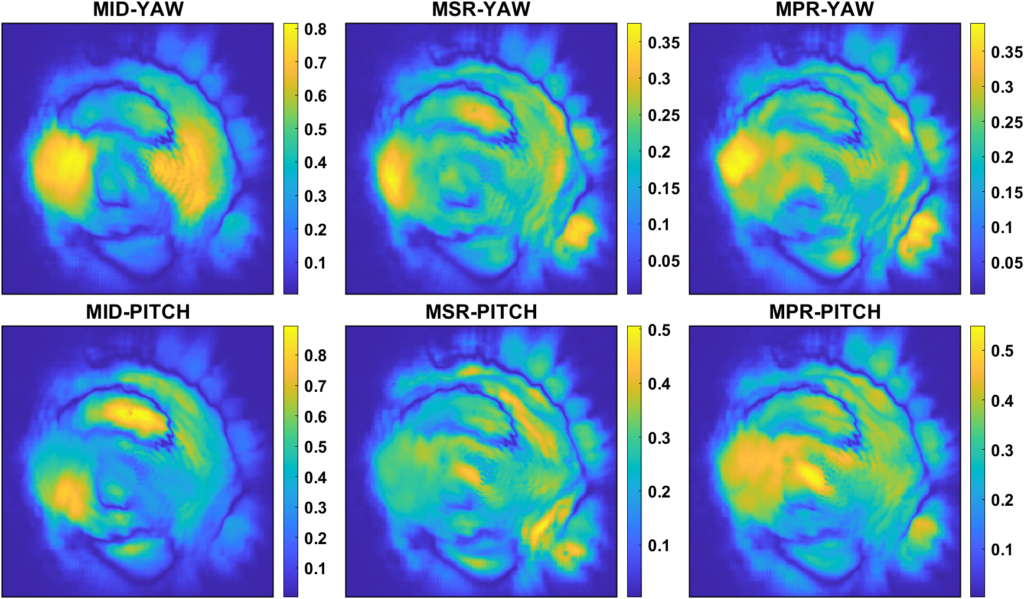}   
       \caption{Coherence maps show the complex coupling of critical alignment degrees of freedom to the interferometric darkport. Coupling is estimated using the weighted coherence in the 0.1-4Hz band. The images from left to right depict the Michelson differential arm motion and the motion from the signal recycling and power recycling mirrors.}
        
    \label{fig:coherence_map}
\end{figure}

\noindent Figure \ref{fig:coherence_map} reveals the complex nature in which the multiple optics imprint their state of alignment on the interferometric darkport. The map is constructed by measuring coherence between the pixel-wise darkport intensity fluctuations and the temporal variation in different alignment error signals. $\mathrm{C_{xy}}$, the metric used for computing the coupling is the magnitude-squared coherence in the 0.1-4 Hz band weighted by the average logarithmic error-point spectra,

\begin{equation}
\mathrm{
C_{xy} = \frac{ \displaystyle \int_{f_{1}}^{f_{2}}\log_{10}(<y(f)^{asd}_{norm}>) \cdot \frac{|P_{xy}(f)|^{2}}{P_{xx}(f)\,P_{yy}(f)}}{\displaystyle \int_{f_{1}}^{f_{2}}\log_{10}(<y(f)^{asd}_{norm}>)}.}
\end{equation}

The high coherence and peculiar spatial spread confirm that darkport contains a treasure trove of information about the IFO. Apart from being a helpful detector characterization tool, coherence maps can be used to identify potential signals that a trained neural network can recover. We find a couple of interesting observations. Angular misalignment causes coupling of the light field into the first order mode, which is well captured by the appearance of (1,0) and (0,1) Hermite Gaussian mode patterns and is most prominent for the Michelson optics. We see more complex and radially extended structures for Signal and Power recycling optics, which could indicate higher-order spatial modes. For all three optics, coherence from pitch seems to be higher than the yaw degree of freedom. This difference is expected since the angle-to-length coupling in suspended optics is more likely to happen via tilt or pitch than the yaw motion.

\subsection{Neural Sensor Architecture}

\begin{table*}[]
    \begin{tabular}{cc|c|c|c|c|}
        \hline
        \multicolumn{1}{|c|}{\multirow{2}{*}{\textbf{Model}}} & \multicolumn{1}{c|}{\textbf{Pre-Trained}} & \multicolumn{1}{c|}{\textbf{Re-Trained}} & \multicolumn{3}{c|}{\textbf{After Quantization}}             \\ \cline{4-6}
        \multicolumn{1}{|c|}{}                                  & \textbf{RMSE} & \textbf{RMSE}                                & {\begin{tabular}[c]{@{}c@{}}Memory \\ Compression\end{tabular}} & {\begin{tabular}[c]{@{}c@{}}Frame rate  \\ Increase\end{tabular}} & {\begin{tabular}[c]{@{}c@{}}RMSE \\ Degradation\end{tabular}}                                    \\ \hline
        \multicolumn{1}{|c|}{Squeezenet-LSTM} & 0.135                                          & 0.118           & $\times$ 3.96           & $\times$ 5.9           & 8.3 \%                            \\ \hline
        \multicolumn{1}{|c|}{Googlenet-LSTM} & 0.122                                          & 0.105           &  $\times$ 3.98          & $\times$ 4.6           & 4.3 \%                            \\ \hline
        \multicolumn{1}{|c|}{InceptionResnetV2-LSTM} & 0.132                                          & 0.096           &  $\times$ 3.98           &  $\times$ 4.1          & 21.9 \%                            \\ \hline        
    \end{tabular}
     \caption{Sensor prediction errors for the three neural architectures. We compare the RMSE for the trained LSTM layers with pre-trained CNN and the fine-tuned CNN-LSTM network. The last three columns provide the results after quantizing the combined network from single-bit floating point to INT8 precision.}
    \label{tab:NN_COMPARE}    
\end{table*}

\noindent The neural alignment sensing we intend to do can be formulated as an image-to-time-series regression problem. We choose a CNN-LSTM architecture for this task for a few reasons. While 2-D convolutional neural nets (CNN) are well suited for analyzing image data with complex spatial representation, long short-term memory networks (LSTM) \cite{HochSchm97} excel at temporal modeling and sequence prediction. LSTMs, a specific form of recurrent neural networks, use a memory cell that selectively controls the flow of information using input, output, and forgets gates. They also had limited success at linear system identification tasks, with results comparable to traditional transfer function estimation \cite{LSTM_sysID}. The ability to learn representations in both space and time thus makes the combined deep recurrent convolutional model effective at activity recognition from streaming video data and makes it a good candidate for capturing the underlying system dynamics \cite{donahue2015long,CLDNN}. 

\par Our design choice for the CNN architecture is based on transfer learning \cite{bozinovski1976influence,pratt1991direct} where the initial layers of pre-trained networks, fine-tuned to extract spatial information at different scales by training on standardized datasets, are reused for a newer task. Transfer learning alleviates the need for training networks from scratch and is useful when the data is limited in size.
In particular, we focus on inception-based networks where spatial filters of different scales are convolved in parallel, thus processing information at bigger scales and finer resolution. These networks represent a synergy between classical computer vision and deep architectures and have previously successfully recovered all the GW events listed in the GWTC-1 transients catalog \cite{PhysRevD.104.064051}. We use three reference architectures, namely squeezenet \cite{iandola2016squeezenet}, googlenet \cite{googlenet}, and inceptionResnetV2 \cite{szegedy2017inception} and compare the respective trade-offs. We chose squeezenet as a comparatively lightweight network with 18 layers, making it suitable for embedded devices and low latency inference. The 164 layers-deep inceptionResnetV2  is among the largest pre-trained networks and provides high classification accuracy on several benchmark datasets. Googlenet is often a good choice when we must balance network size and accuracy.

\subsection{Training Strategy}

\begin{figure*}[!htb]
\centering
\includegraphics[width=0.95\linewidth]{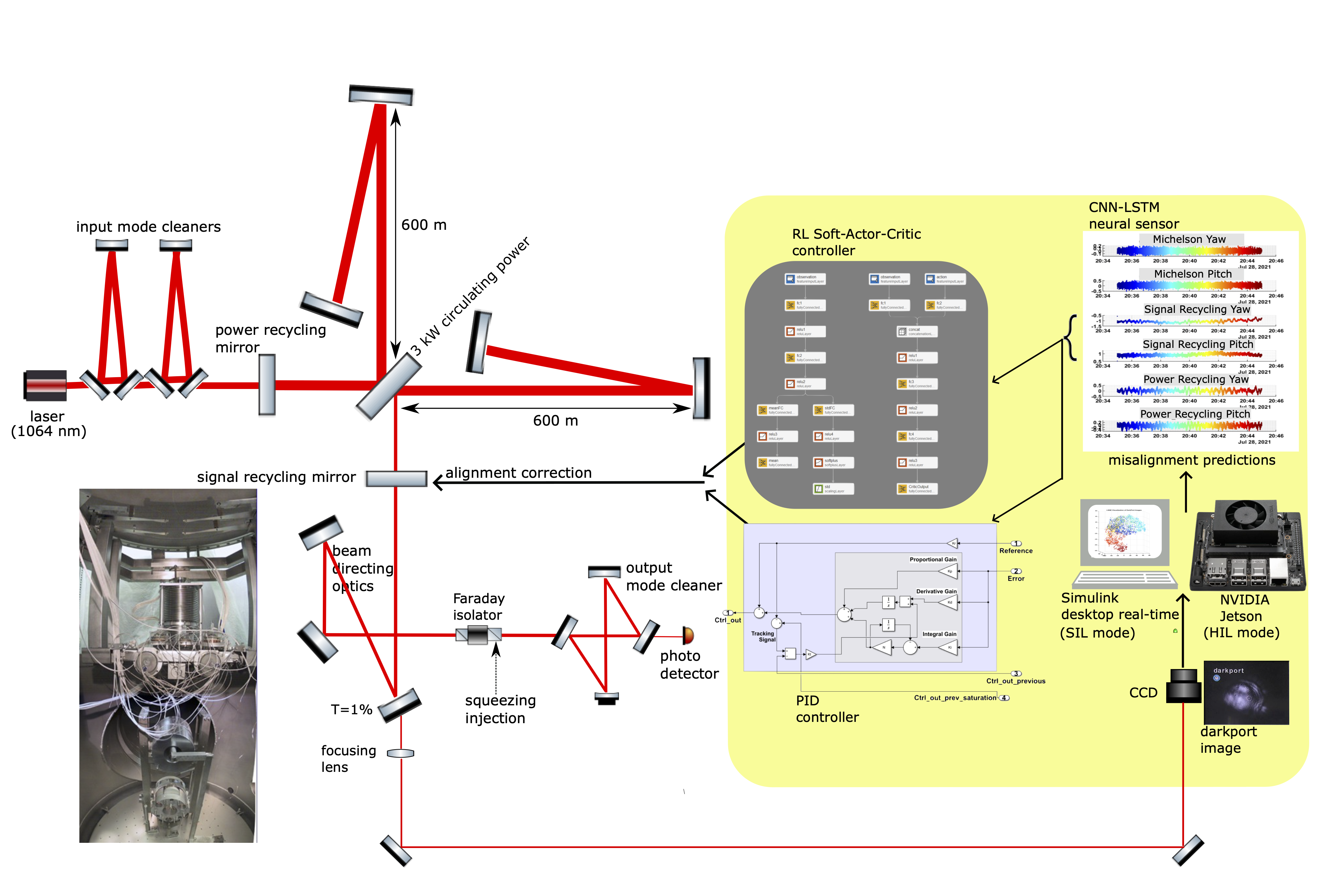}
\caption{Simplified optical layout of GEO\,600 highlighting the AI-based alignment sensing and control scheme. The CCD captures 2D images of the beam that exit the darkport through the 1\% transmission port of the beam-directing optic. The CNN-LSTM neural network simultaneously extracts the pitch and yaw degrees of freedom for the Michelson, signal recycling, and power recycling mirrors. Tuned PID controllers and soft-actor-critic-based reinforcement learning agents process this information and correct the low-frequency drifts of the signal recycling mirror, thus improving the astrophysical sensitivity.}
\label{fig:GEO_layout}
\end{figure*}

\noindent Our aim with the CNN-LSTM model is to train the network on sufficient darkport images and the corresponding DWS alignment signals from a well-tuned interferometer configuration and predict the new error points whenever the detector gets into a misaligned state. If the model is well-trained, it should be able to predict the current loop offset value affected by drifts, and then either a human or a controller (either classical PID or reinforcement learning based) can set it to the last known "good" state. Figure \ref{fig:GEO_layout} gives the corresponding sensing and control schematic.

Training deep networks is, in general, a time-consuming process, and additionally, we need also to find the right hyper-parameters to maximize the learning process. We adopt a strategy where we start with squeezenet, cut the network just before the final fully connected layers and add the LSTM layers with ten hidden units to its output. We then freeze the weights of squeezenet layers and let the LSTM layers learn while the combined network is trained to predict the alignment error points from the recorded darkport images. This configuration makes it easier to determine parameters like gradient decay rate, LSTM hidden units, and learn rate using minimal computational resources. In the second stage of training, we retrain the entire network comprising the pre-trained CNN and newly trained LSTM and fine-tune the network weights and biases. This stage of training is carried out on a dedicated A100 GPU cluster. We repeat the process for the other two pre-trained networks. Table \ref{tab:NN_COMPARE} compares the root mean squared error between the neural sensor predictions and the actual alignment signals for the previously mentioned six degrees of freedom.

\begin{figure}[!htb]
    \centering
    \includegraphics[width=\linewidth]{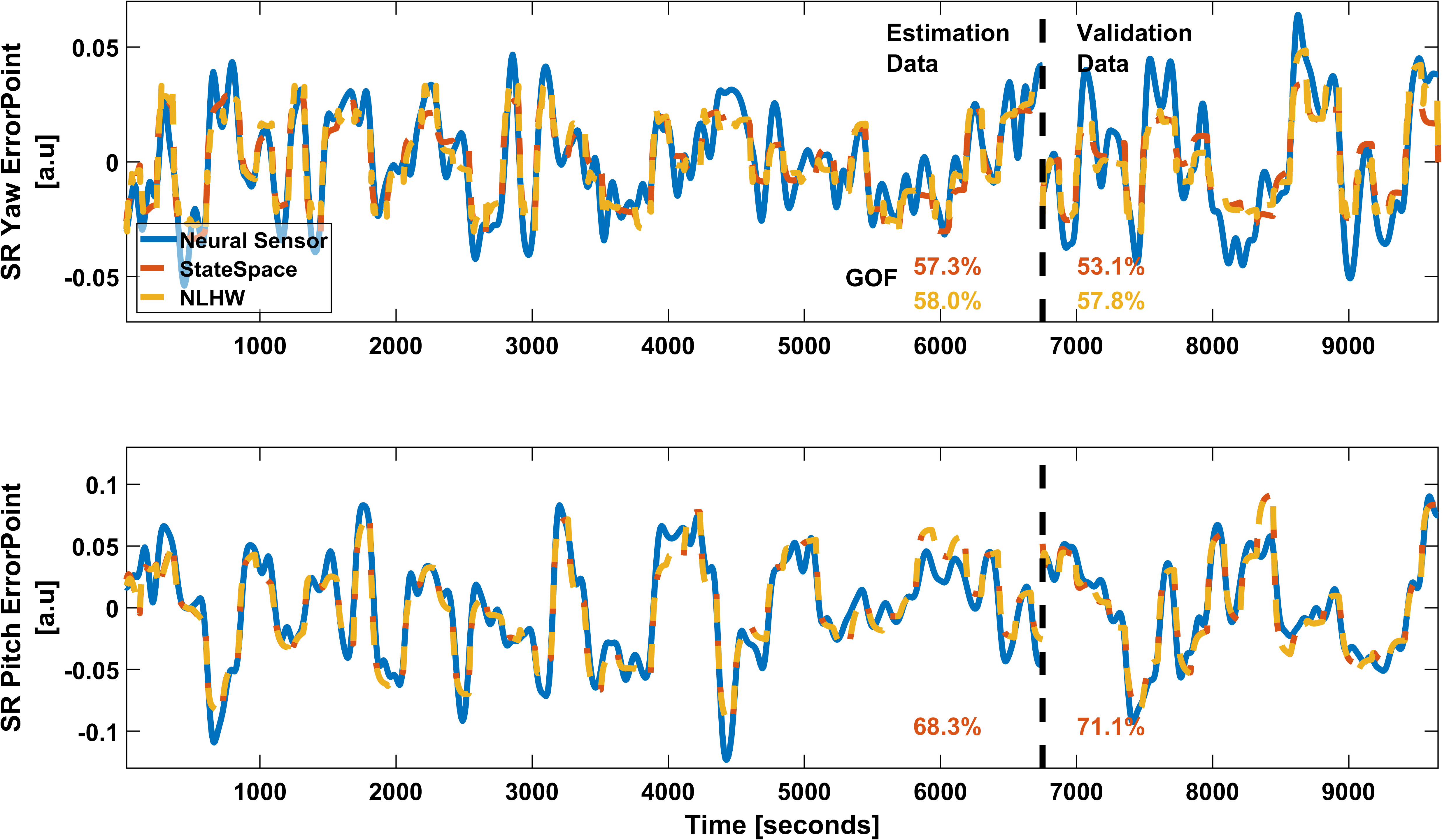}
    \caption{Comparison of neural sensor inferred system dynamics with reduced order models. The identified state space model is used to design the optimal PID controller and train the reinforcement learning agent.}
    \label{fig:statespace_sysid}
\end{figure}

\subsection{Network Quantization}

\noindent Most often, the learnable parameters of neural networks are trained using single-precision floating point data types. However, the limited dynamic range of these parameters makes it possible to cast them as scaled 8-bit integer data types of fixed length. Such quantization can significantly reduce the memory footprint, improve the inference rate, and lower the power consumption \cite{courbariaux2014training,han2015deep}. This step would be crucial when the trained networks are deployed at a large scale in GW detectors using embedded devices like FPGAs, ASICs, or GPU-accelerated EDGE devices for real-time processing. We use a training data set to calibrate the dynamic range of the weights and biases of the convolutional and fully connected layers and the activations in all the layers. Using a separate validation dataset, we quantize to the right data type (single-bit floating point or INT8), ensuring to cover the range, avoiding overflows but ignoring potential underflows. Table \ref{tab:NN_COMPARE} gives the memory reduction and the improved processing speed, measured in terms of frames per second, and the decrease in accuracy for the three quantized network architectures.

We select fine-tuned Googlenet-LSTM for the rest of our analysis as it provides a decent trade-off among metrics like time for training,  prediction accuracy, and the real-time inference rate. 

\section{Model-Based Controller Design} \label{Controller}

\noindent After the neural sensor is built as described above, we require a suitable controller to close the loop. Designing such controllers with the actual interferometer-in-loop is not encouraged. Doing so reduces the observation time and can lead to undesired behavior like oscillations in the system that could take a long time to settle down. For example, an RL agent, in its search for the optimal policy, can intentionally carry out random action sequences as it tries a strike a balance between exploration and exploitation. We hence follow a model-based \cite{kapurch2010nasa} approach and design a controller that can utilize the signals from the neural sensor. The original high-fidelity instrument response is approximated by a reduced-order model that sufficiently captures the dominant system dynamics relevant to the controller design. The response of the interferometer is analyzed by randomly perturbing the setpoints in SR pitch and yaw degree of freedom that covers the actuation range of the existing controller. This perturbation leads to variation in the darkport images and is processed by the CNN-LSTM neural sensor. System identification, mapping setpoints to neural predictions, is then carried out using subspace-based state-space modeling \cite{van1994n4sid}. State space offers superior performance over transfer function models due to the ability to include a noise model. During the fitting process, the model order is varied over a reasonable range, and the one with the lowest Hankel singular value \cite{boyd1994linear} is selected. Selecting such lower values helps retain the larger energy states, making it possible to have a reduced-order model that preserves the majority of the system characteristics. The identified model is further refined using the prediction-error minimization, where the weighted norm of the difference between the measurement and the model's predicted output is minimized \cite{ljung1998system}. We looked at further improvements by adding input and output non-linearities to the identified state space model, resulting in a non-linear Hammerstein-Wiener (NLHW) model. Figure  \ref{fig:statespace_sysid} compares both the models on estimation and validation data, where the goodness of fit is given as,

\begin{equation}
    \mathrm{GOF} = 100\,\Big(1- \frac{\|y_{model}-y_{meas}\|}{\|y_{meas}-y_{meas}^{mean}\|}\Big)\,.
\end{equation}

The NLHW model, with a sigmoid network function representing the non-linear mapping, only provides a modest improvement in one degree of freedom. Hence, we select the state space model for the rest of the analysis.

\subsection{Classical PID-based Controller}

\noindent Proportional-Integral-Derivative (PID) controllers are among the most widely used classical controllers for linear-time-invariant (LTI) systems. They are easy to tune, depend only on the error signal, and are less susceptible to plant variations. They can be designed to ensure closed-loop stability of the plant \cite{franklin2002feedback} and also serve as a benchmark while evaluating the performance of the RL-based solutions described in the next section. These linear controllers, however, need to be separately tuned for each DOF. We use the plant model identified in the previous section and automate the tuning focusing on reference tracking. Tunable parameters are obtained using H-infinity synthesis by optimizing across the target bandwidth, performance, and robustness requirements \cite{bruinsma1990fast, Apkarian2006}. However, the presence of an actuator with a limited range introduces non-linearities and often leads to the well-known integral windup \cite{aastrom2006advanced}. We overcome this using additional anti-windup circuity built using a tracking signal and a reference feed-forward. The output for the controller with an error signal $\mathrm{e(t)}$, depicted in Figure  \ref{fig:GEO_layout}, is given by, 

\begin{multline}
    u(t) = \mathrm{K_{p}}\,e(t) \, + \, \mathrm{K_{d}}\,\frac{d\,e(t)}{dt}\,+\, \mathrm{K_{r}}\,r(t) \\ \,+\, \int \Bigg[ \mathrm{K_{i}}\,e(t) \,+\,  \mathrm{K_{t}}\,u_{s}(t)\,-\,\mathrm{K_{t}\,K_{r}}\,r(t) \,-\, \mathrm{K_{t}}\,u(t) \Bigg]\,dt.
\end{multline}

\noindent where $\mathrm{K_{p}}$, 
$\mathrm{K_{i}}$, 
$\mathrm{K_{d}}$ 
are the usual PID gain coefficients,
$\mathrm{K_{r}}$ controls the reference $\mathrm{r(t)}$ feedforward, while the tracking coefficient $\mathrm{K_{t}}$ 
and the saturated output $\mathrm{u_{s}}$ are part of the modified integral term.

\subsection{Deep Reinforced Controller}

\noindent Reinforcement learning (RL) is an experience-based learning framework that eliminates the need for supervision and subject expertise and attempts to learn to carry out a task based purely on its interaction with the system \cite{sutton2018reinforcement}. The notion of a traditional controller is replaced here by an RL agent consisting of a deep neural network and a policy-updating algorithm. The former provides high-capacity representations that are easy to generalize, while the latter offers a mathematical formalism for decision-making and optimal control. During the training process, the agent observes the system's current state, interacts with the environment, and considers the new states and the reward, an immediate measure of the goodness or badness of the current action. The agent then tries to learn the optimal policy, or the mapping between states and actions, to maximize the discounted cumulative long-term reward.

\subsubsection{Soft-Actor-Critic Algorithm}

\noindent Traditional RL algorithms were thought to be unstable and unpredictable, making them sensitive to hyper-parameters and initial conditions. One way to overcome this scenario is to cast RL and the optimal control as a probabilistic inference problem. Soft actor-critic (SAC) consists of a set of algorithms \cite{haarnoja2018soft} that utilizes the traditional actor-critic methods \cite{sutton1999policy,mnih2016asynchronous,schulman2015high,gu2016q} but ensures maximization of the entropy of the learned policy. The action-value function (or the Q-function), which evaluates the quality of the agent's actions, is determined using a pair of critic networks, thus minimizing the over-estimation bias. They are trained using the Bellman equation, which involves an iterative update of the value function whenever a state-action pair is traversed by the agent and is given by,

\begin{multline}
\mathrm{Q^{new}(S,A) = Q^{prev}(S,A) \;+}  \\ \mathrm{\alpha \Bigg[  \Bigg( R(S,A) + \gamma \; \underset{A'}{max}\, Q(S',A') \Bigg)  - Q^{prev}(S,A)   \Bigg]}
\end{multline}

\begin{table*}
\hrule
\centering
\rule{\textwidth}{\heavyrulewidth}
\vspace{-\baselineskip}
\begin{flalign}
&
\text{Cost} 
& 
   \mathrm{\sum_{j=1}^{\tau} (S_{j}-S^{ref}_{j})^{T} \, \mathbb{Q}_{j} \, (S_{j}-S^{ref}_{j})} \; + \; \mathrm{(A_{j}-A^{prev}_{j})^{T} \, \mathbb{R}_{j} \, (A_{j}-A^{prev}_{j})\,}  
&&&
\label{Cost_func}
\end{flalign}
\vspace{-\baselineskip}
\vspace{-\baselineskip}
\begin{flalign}
&
\text{Penalty}
&
\hspace{7em}
    \mathrm{W_{y}\,\Bigg( (S_{j}-S^{min}) ^{2} + (S_{k}-S^{max}) ^{2} \Bigg)} + \mathrm{W_{mvrate} \, \Bigg( (\dot{A}_{l}-\dot{A}^{max})^{2} +  (\dot{A}_{m}-\dot{A}^{min})^{2} \Bigg)\,}
&&&
\label{Penalty_eqn}
\end{flalign}
\vspace{-\baselineskip}
\vspace{-\baselineskip}
\begin{flalign}
&
\text{}
&
\hspace{18em}
 \forall \mathrm{\Bigg( \; S_{j}<S^{min}, \; S_{k}>S^{max}, \; \dot{A}_{l}<\dot{A}^{min},\; \dot{A}_{m}>\dot{A}^{max}  \Bigg) \, } 
&&&
\label{Penalty_condition_eqn}
\end{flalign}
\vspace{-\baselineskip}
\vspace{-\baselineskip}
\begin{flalign}
&
\text{Boost}
&
\hspace{-1em}
 \mathrm{ 10\,\sum_{j=1}^{\tau} (3\,| S_{j}-S^{ref}_{j}|<0.02)^{2} } \; + \; \mathrm{10\,\sum_{j=1}^{\tau} (6\,| S_{j}-S^{ref}_{j}|<0.005)^{2}\,}
&&&
\label{Boost_eqn}
\end{flalign}
\vspace{-\baselineskip}
\vspace{-\baselineskip}
\begin{flalign}
&
\text{Reward}
&
\hspace{-14em}
   \; -( \mathrm{Cost}\;+ \mathrm{Penalty}) \;+\;  \mathrm{Boost}\,
&&&
\label{Reward_eqn}
\end{flalign}
\vspace{-\baselineskip}
\vspace{-\baselineskip}
\rule{\textwidth}{\heavyrulewidth}
\hrule
\caption{Reward function equations used to train the reinforcement learning agent. The continuous part is built using the linear quadratic regulator cost function. Discrete terms, penalty and boost, are added to penalize the violation of boundary constraints and emulate final state constraints.}
\label{tab:RL_REWARD}
\end{table*}

\noindent where $\mathrm{\gamma}$ is the discount factor for future rewards and  $\mathrm{\alpha}$ controls the value update learning rate for a given state-action (S, A) pair. The actor network representing the policy $\mathrm{\pi}$, is trained using the gradient of the expected return concerning the actions, which is computed using the critic network. By learning a probabilistic regularized "soft" policy trained to maximize both value and policy entropy, 

\begin{equation}
\mathrm{ \underset{\pi}{max} \; \mathbb{E}_{\pi}  \Big[  Q(S,A) - log \, \pi (A|S)  \Big]  \,,}
\end{equation}

\noindent the agent learns a wide range of behaviors, including stochastic or deterministic behaviors. A comparatively faster learning rate, lower sensitivity to hyper-parameters, ability to reuse fast experience, and a balanced trade-off between exploration and exploitation make SAC a good candidate for real-world control problems.

    An ideal reward function should guide the agent to the optimal policy. However, creating a suitable reward function is the most critical task in RL training. One goal of this work is to assess the practicality of this approach in designing controllers suitable for GW detectors and probe if a set of general guiding principles can help design a reward that leads the agent to the optimal policy. Table \ref{tab:RL_REWARD}  lists the components of the reward function used to train our RL agent.
    We draw cues from optimal control theory, which aims to operate dynamical systems with minimal controller effort. The continuous portion of the reward can be derived from the corresponding linear-quadratic-regulator (LQR) cost function. For LTI systems with a quadratic cost function, LQR provides the optimal gain matrix for state feedback control by solving the Riccati equation of the state-space model. The corresponding cost that drives the state close to the reference with minimal actuator effort is expressed in terms of both the current and reference state $\mathrm{(S_{j}, S_{j}^{ref})}$, and the current and previous actuator values $\mathrm{(A_{j}, A_{j}^{prev})}$, with $\mathbb{Q}_{j}$  and $\mathbb{R}_{j}$ being the respective weight matrices. Such continuous rewards encourage convergence but are prone to local minima and can lead to longer training periods. Adding discrete elements that penalize or encourage the agent increases the probability of finding better states. However, the non-smooth nature of the resulting loss function can affect the convergence. We discourage boundary constraint violations from the agent by including a discrete penalty term, where behaviors that drive the states close to the limits $\mathrm{(S^{min}, S^{max})}$ or increase controller velocity beyond a threshold value $\mathrm{(\dot{A}^{min},\dot{A}^{max})}$ are penalized, with $\mathrm{W_{y}}$ and $\mathrm{W_{mvrate}}$ being the associated weight matrices. We also observe the benefit of including a discrete positive reward when the state is driven close to the reference.

\begin{table}[!htb]
\begin{tabular}{ | p{4.5em} | p{4.5em} | p{6.5em} | p{4.5em} | }
  \hline \centering  PID Active & \centering RL agent Active & \centering RL agent Type &  Relative average reward \\
  \hline \centering  Yes & \centering No  & \centering  -                  & \;\;\; 0.95 \\
  \hline \centering  No  & \centering Yes & \centering Single, suboptimal & \;\;\; 0.80 \\
  \hline \centering  0.7 & \centering 0.3 & \centering Single, suboptimal & \;\;\; 0.96 \\ 
  \hline \centering  0.5 & \centering 0.5 & \centering Single, suboptimal & \;\;\; 0.96\\
  \hline \centering 0.5  & \centering 0.5 & \centering Ensemble, optimal   & \;\;\; 0.96  \\
  \hline \centering 0.3  & \centering 0.7 & \centering Single, suboptimal & \;\;\; 0.94  \\
  \hline \centering 0.3  & \centering 0.7 & \centering Ensemble, optimal   & \;\;\; 0.97 \\         
  \hline \centering No  &  \centering Yes & \centering Ensemble, optimal   & 
  \;\;\;\;\; 1 \\                     
  \hline
\end{tabular}
      \caption{Comparison of different controllers for alignment reference tracking of the SR mirror.}
    \label{tab:rl_pid_comparison}
\end{table}

\begin{figure*}[!htb]
    \centering
    \includegraphics[width=\linewidth]{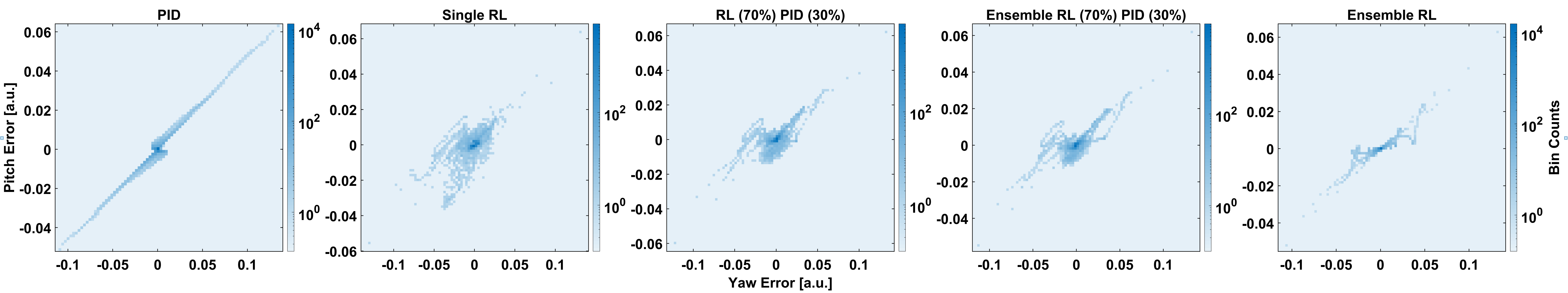}
    \caption{Bias and variance in 2-DOF reference tracking for the controllers mentioned in Table \ref{tab:rl_pid_comparison}}
    \label{fig:bin_scatter}
\end{figure*}

\par The random initialization of network weights and entropy maximization objective associated with the optimal policy learning make the overall convergence rate moderately sensitive to individual simulation runs. Hence, we carry out ten training trials for each network configuration. One usual design decision is to choose between a deeper or wider network. In supervised learning tasks, issues with vanishing gradients make it harder to train deeper networks and are usually overcome using residual connections. However, in the case of the RL agent, the difficulty in training deeper networks arise from the sharpness of the loss surface curvatures, making them more susceptible to the choice of hyperparameters. Recent studies \cite{ota2021training} prefer the wider networks as they have nearly convex loss surfaces. We observe similar performance improvement with increased network width (see Appendix).

\begin{figure*}[!htb]
     \centering  \includegraphics[width=0.95\textwidth]{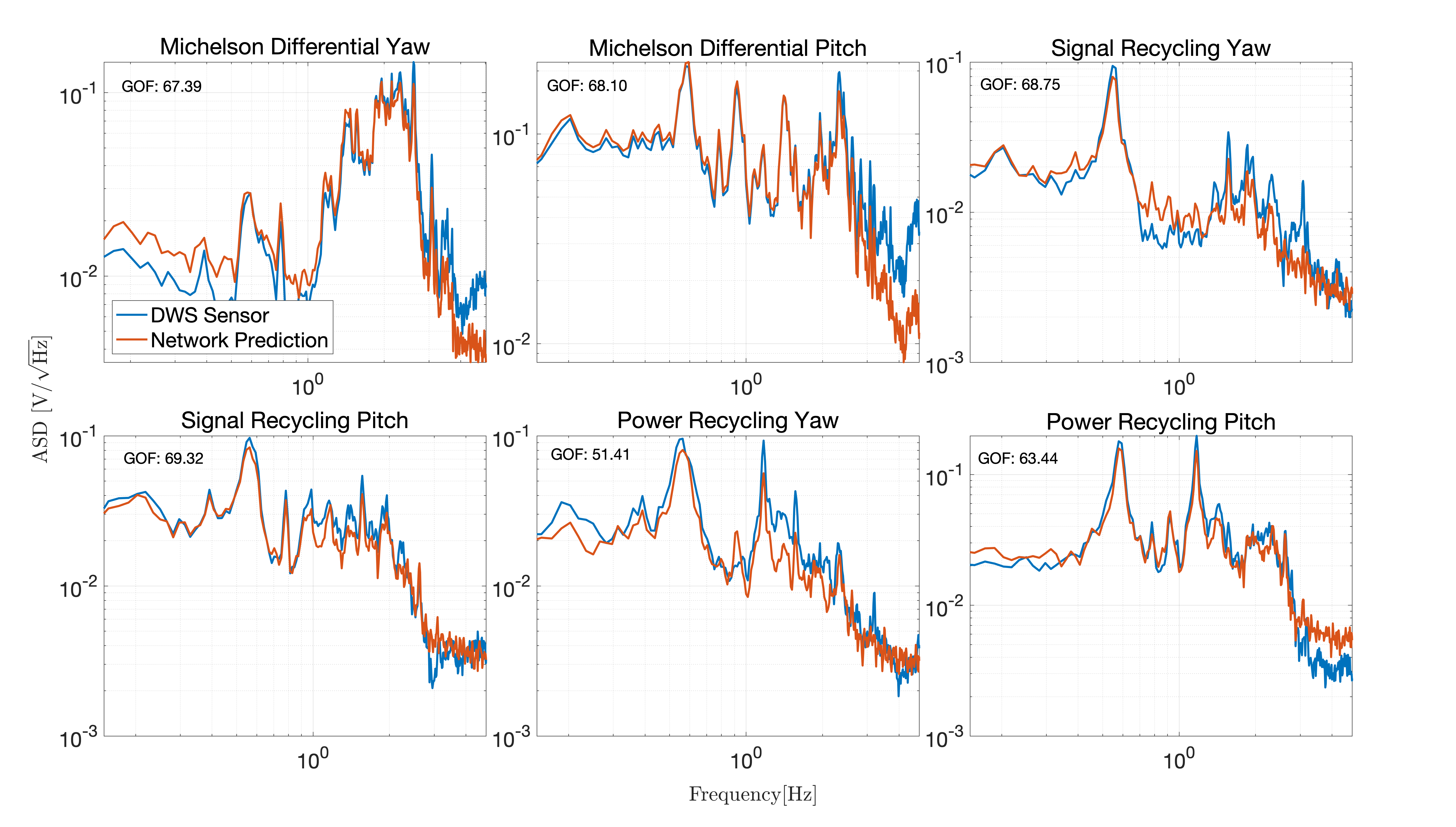}
 \caption{Comparison of neural network alignment predictions for the key optics with the measurements from the differential wavefront sensor.}    
     \label{fig:reTrainFreqPRED}
\end{figure*}

\subsection{Multi-Agent control}

\noindent Ideally, the designed controllers should be less susceptible to the uncertainties associated with the modeled environment and our limited knowledge of the optimal reward. One way to achieve robustness is by blending in control signals and leveraging the positive aspects of each, such as the low integral error from PID and the faster response of RL. Ensemble learning \cite{wiering2008ensemble} is another option, where the top-performing agents across the multiple simulations are combined to form the optimal signal by averaging the maximum likelihood action suggested by each. We report the findings from simulating probable controller combinations in Table \ref{tab:rl_pid_comparison}. It includes a tuned PID for each DOF, a single RL agent with an average performance, an ensemble of optimally performing RL agents, and a few combinations where the signals are blended. The ensemble learner achieves the best performance measured in terms of the recovered average reward, where the setpoints are randomly perturbed across the actuation range. The corresponding bias and variance associated with the 2-DOF reference tracking for each controller configuration are shown in Figure \ref{fig:bin_scatter}.

\section{Results}\label{RESULTS}

\begin{figure*}[!htb]
    \centering
    \includegraphics[width=0.75\linewidth]{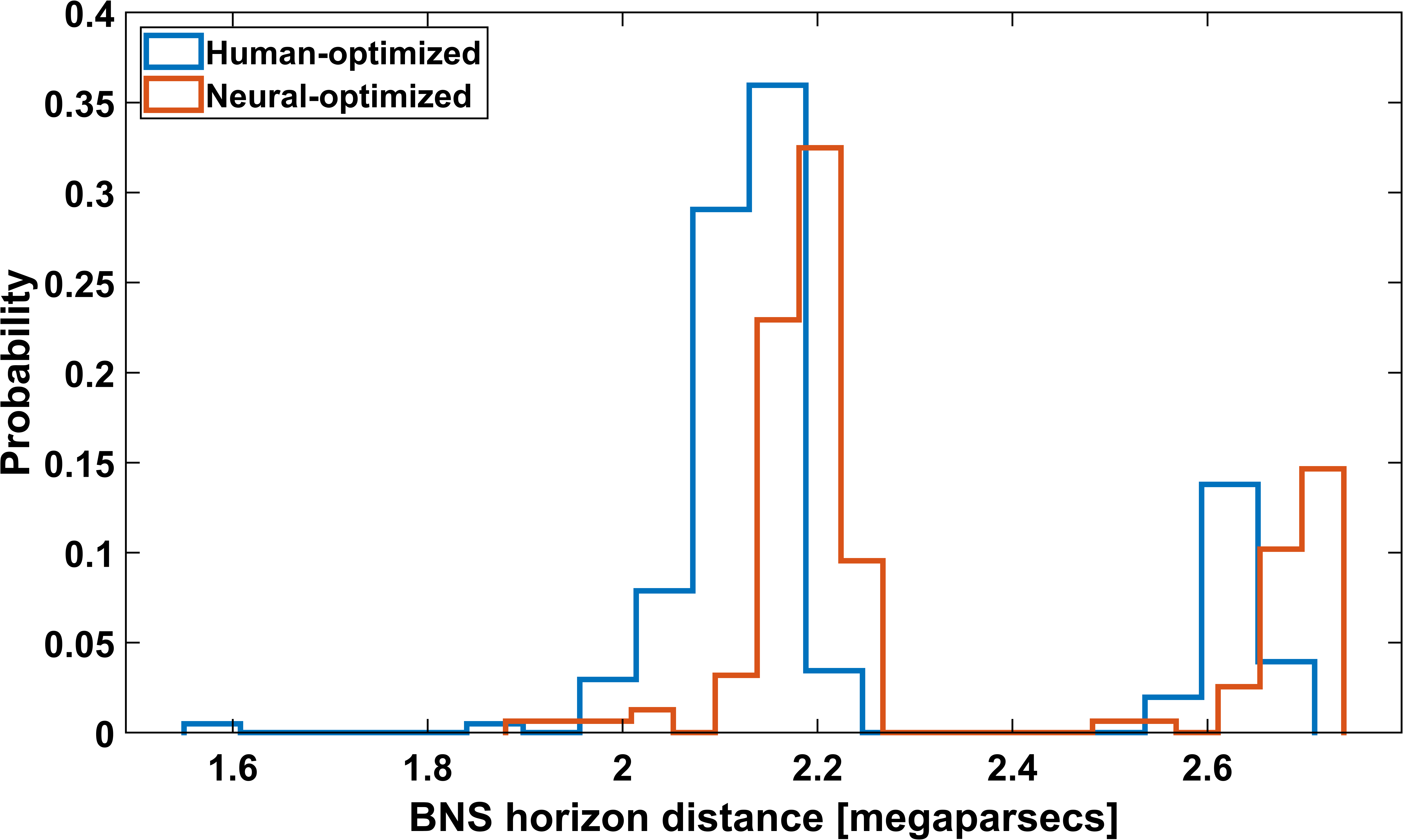}
    \caption{Astrophysical sensitivity comparison in terms of horizon distance for coalescing neutron star binaries of 1.4 solar mass each. The blue trace is when experienced commissioners fine-tune the interferometer, while the orange trace provides the results using the deep reinforcement learning agent. The two peaks represent the scenario with and without applying squeezed states of light.}
    \label{fig:hsensmon}
\end{figure*}

\noindent In Figure  \ref{fig:reTrainFreqPRED}, we present the alignment predictions made by the retrained InceptionResnetV2-LSTM neural sensor and compare them with the actual measurements for the six alignment DOFs (also see Table \ref{tab:NN_COMPARE}). These signals are generated from unseen darkport image sequences, and we observe a good match in both time and frequency domains with true error points (see Table \ref{tab:NN_COMPARE}). We finally close the AI-based alignment loop by deploying the neural sensor and the ensemble RL agent in real-time, as shown in Figure \ref{fig:GEO_layout}, and accessing the resulting performance. 

The ultimate metric to access the impact of the neural scheme is to analyze the GW strain curve and estimate the improvements to astrophysical sensitivity. The non-stationary nature of the noises influencing the detector, primarily the ambient seismic noise, often makes comparing different time segments difficult. To address this, we measure several pairs of segments approximately thirty minutes long, each with the SR mirror optimized manually and with the AI-based controller. Figure \ref{fig:hsensmon} compares this sensitivity in terms of the farthest luminosity distance for optimally orientated and located coalescing neutron star binaries (1.4 solar mass each), detectable with a matched filter SNR of eight or above. The blue trace shows the horizon distance when experienced commissioners fine-tune the interferometer by inspecting the darkport images, while the orange trace depicts the results obtained using the deep reinforcement learning agent. The first peak is when the interferometer is shot noise limited at high frequencies, a state similar to the training data. The second peak is when the interferometer is operating at a modified state with the injection of squeezed states of light. In both cases, we observe the neural-optimized segments to outperform the manual tuning from the experienced commissioners.

\section{Conclusions and Outlook}
\noindent With GW detectors becoming more complex with each generation, AI-assisted autonomous sensing, and control could play a major role in the operation of interferometers, including automated alignment and multi-cavity locking. We developed a deep neural network scheme to extract meaningful information about the state of the interferometer and reconstructed the alignment error signals using the data from GEO\,600 observatory. We implemented a control loop using this neural sensor and achieved drift control of the signal recycling mirror using deep reinforcement learning, improving overall sensitivity. As far as we know, this work is the first of its kind where machine learning-based control applied to a kilometer-scale GW interferometer resulted in astrophysical sensitivity improvement. Radio-frequency sidebands from the existing auto-alignment scheme elevate the shot noise at kilohertz frequencies and affect the strain signal. Fully replacing this method with a higher bandwidth version of the neural scheme presented here is an interesting possibility that is part of future work.

\par We followed a divide-and-conquer approach, deploying different neural architectures and multiple learning strategies for sensing and control. End-to-end learning using a single transformer-based architecture with self-attention \cite{vaswani2017attention} could lead to a better flow of gradients and improved predictions. 
Expanding the RL-controller's policy to include a diverse set of tasks would also be desirable if we intend to control multiple sub-systems. DeepMind's Gato \cite{reed2022generalist} and Robotics Transformer (RT-1) from Google Brain \cite{brohan2022rt} have recently demonstrated the most promising strides towards artificial general intelligence, enabling multitask learning using a context-based generalized policy. The applicability of such frameworks that combine transformer models with reinforcement learning strategies looks promising for current and future generation GW observatories, and our work is the first step in that direction.

\section{Training  Resources}
\noindent CNN-LSTM neural sensor was built and trained using MATLAB R2022a, while the RL-SAC MIMO controller was set up and trained using the corresponding Simulink modeling environment. GPU training was carried out at the Caltech LIGO cluster (AMD EPYC 7763 64-Core, 256 GB RAM) using the NVIDIA A100-80 GB GPU. The neural network quantization from single-bit floating point to INT8 data type was carried out for SIL and HIL mode, respectively, using the Intel-MKL deep learning library and NVIDIA Jetson Xavier NX.

\section{Acknowledgements}
\noindent NM thanks Thomas K\"unzel, Franziska Albers, and MathWorks for their technical support. NM thanks Kong Chun-Wei, University of Michigan, for his insights about training deep RL controllers. We thank the LIGO control systems working group for their valuable suggestions. The authors are grateful for computational resources provided by the LIGO Laboratory and supported by National Science Foundation Grants PHY-0757058 and PHY-0823459. We thank the GEO collaboration for the development and construction of GEO 600. We also extend our thanks to Walter Gra\ss{} for his work keeping the interferometer in a good running state. We are grateful for support from the Science and Technology Facilities Council (STFC), the University of Glasgow in the UK, the Max Planck Society, the Bundesministerium f\"ur Bildung und Forschung (BMBF), the Volkswagen Stiftung, the cluster of excellence QUEST (Centre for Quantum Engineering and Space-Time Research), the International Max Planck Research School (IMPRS), and the State of Niedersachsen in Germany. This paper has LIGO Document Number LIGO-P2200407.

\appendix

\section{RL-agent training}

\begin{figure}[!htb]
    \centering
    \includegraphics[width=0.95\linewidth]{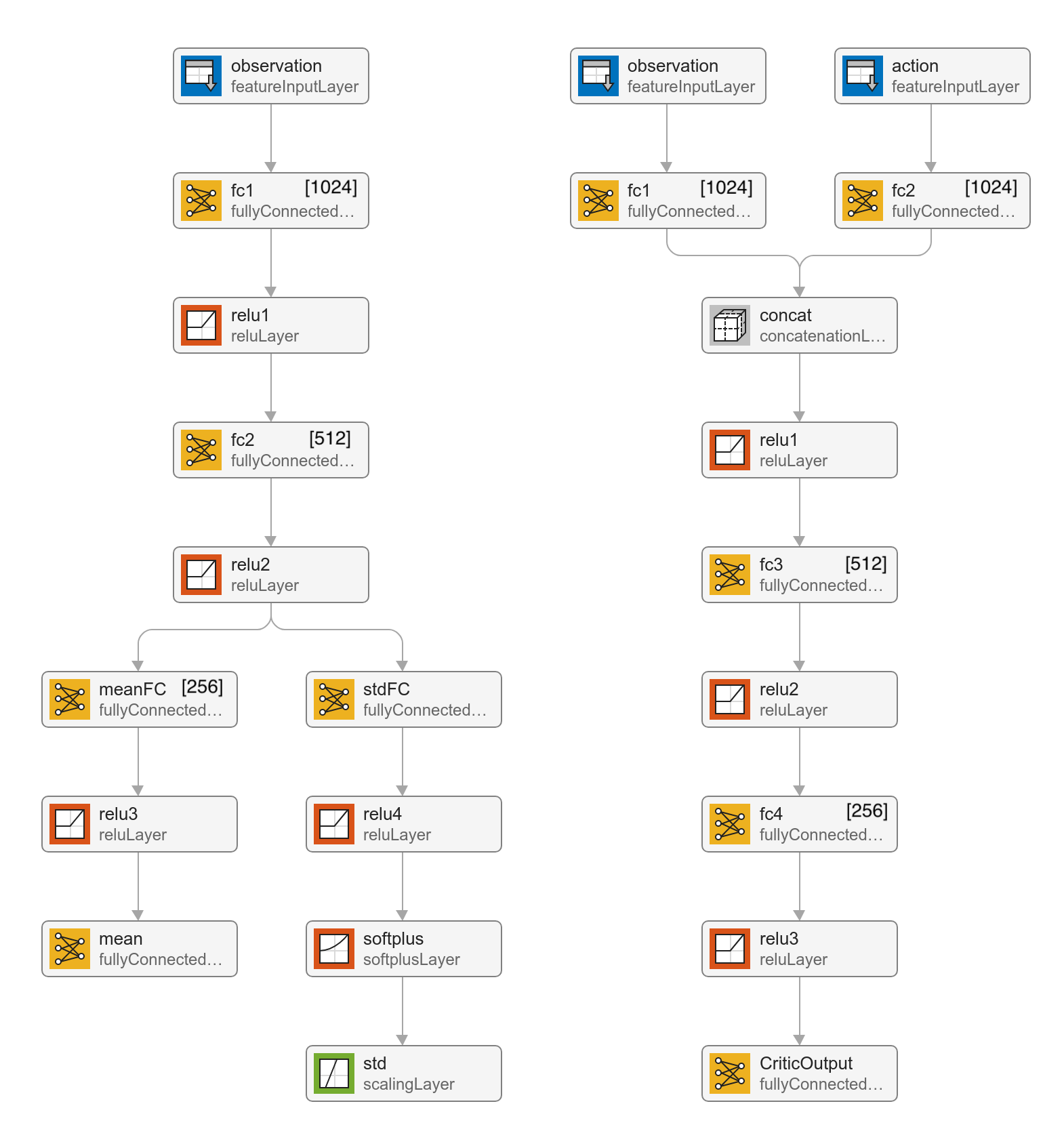}
    \caption{Architecture for the actor and critic networks used to create the SAC agent. The fully connected layers' output size is shown within the square brackets.}
    \label{fig:SAC}
\end{figure}

\begin{figure}[!htb]
    \centering
    \includegraphics[width=0.95\linewidth]{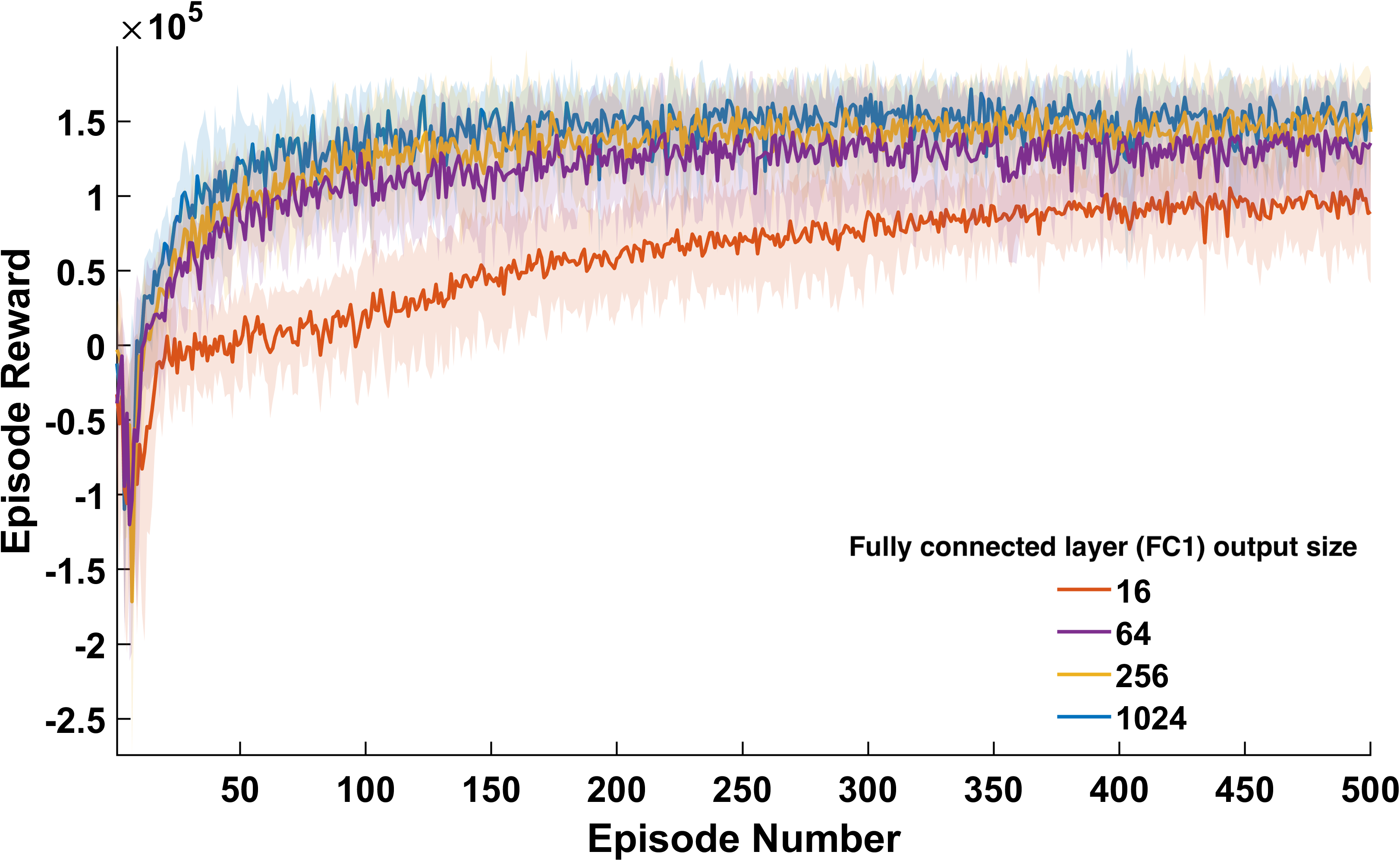}
    \caption{Average episode reward received for RL agents with increasing network size. The output size of the first fully connected (FC1) layer is indicated, and other FC layers are proportionally changed, similar to Figure \ref{fig:SAC}. The shaded region gives the standard deviation errors obtained from ten independent trials.}
    \label{fig:RL-TRAINING}
\end{figure}

\clearpage
\bibliographystyle{unsrt}
\bibliography{References}

\begin{thebibliography}{10}

\bibitem{Grote_2004}
H~Grote, A~Freise, M~Malec, G~Heinzel, B~Willke, H~L\"uck, K~A Strain, J~Hough,
  and K~Danzmann.
\newblock Dual recycling for {GEO} 600.
\newblock {\em Classical and Quantum Gravity}, 21(5):S473--S480, feb 2004.

\bibitem{H_Lueck_2010}
H~Lueck et~al.
\newblock The upgrade of {GEO} 600.
\newblock {\em Journal of Physics: Conference Series}, 228(1):012012, may 2010.

\bibitem{Dooley_2016}
K~L Dooley, J~R Leong, et~al.
\newblock Geo 600 and the geo-hf upgrade program: successes and challenges.
\newblock {\em Classical and Quantum Gravity}, 33(7):075009, mar 2016.

\bibitem{Grote_2010}
H~Grote and (forthe LIGO Scientific~Collaboration).
\newblock The geo 600 status.
\newblock {\em Classical and Quantum Gravity}, 27(8):084003, apr 2010.

\bibitem{10.1093/ptep/ptab018}
T~Akutsu et~al.
\newblock {Overview of KAGRA: Calibration, detector characterization, physical
  environmental monitors, and the geophysics interferometer}.
\newblock {\em Progress of Theoretical and Experimental Physics}, 2021(5), 02
  2021.
\newblock 05A102.

\bibitem{Somiya_2012}
Kentaro Somiya.
\newblock Detector configuration of kagra the japanese cryogenic
  gravitational-wave detector.
\newblock {\em Classical and Quantum Gravity}, 29(12):124007, jun 2012.

\bibitem{ligo2022first}
LIGO~Scientific Collaboration, Virgo Collaboration, KAGRA Collaboration,
  R~Abbott, H~Abe, F~Acernese, K~Ackley, N~Adhikari, RX~Adhikari, VK~Adkins,
  et~al.
\newblock First joint observation by the underground gravitational-wave
  detector kagra with geo 600.
\newblock {\em Progress of Theoretical and Experimental Physics},
  2022(6):063F01, 2022.

\bibitem{Affeldt_2014}
C~Affeldt, K~Danzmann, K~L Dooley, H~Grote, M~Hewitson, S~Hild, J~Hough,
  J~Leong, H~L\"uck, M~Prijatelj, S~Rowan, A~R\"udiger, R~Schilling,
  R~Schnabel, E~Schreiber, B~Sorazu, K~A Strain, H~Vahlbruch, B~Willke,
  W~Winkler, and H~Wittel.
\newblock Advanced techniques in {GEO} 600.
\newblock {\em Classical and Quantum Gravity}, 31(22):224002, nov 2014.

\bibitem{Aasi_2015}
J~Aasi, The LIGO~Scientific Collaboration, et~al.
\newblock Advanced ligo.
\newblock {\em Classical and Quantum Gravity}, 32(7):074001, mar 2015.

\bibitem{PhysRevD.93.112004}
D.~V. Martynov, E.~D. Hall, et~al.
\newblock Sensitivity of the advanced ligo detectors at the beginning of
  gravitational wave astronomy.
\newblock {\em Phys. Rev. D}, 93:112004, Jun 2016.

\bibitem{Acernese_2015}
F~Acernese et~al.
\newblock Advanced virgo: a second-generation interferometric gravitational
  wave detector.
\newblock {\em Classical and Quantum Gravity}, 32(2):024001, dec 2014.

\bibitem{PhysRevLett.110.181101}
H.~Grote, K.~Danzmann, K.~L. Dooley, R.~Schnabel, J.~Slutsky, and H.~Vahlbruch.
\newblock First long-term application of squeezed states of light in a
  gravitational-wave observatory.
\newblock {\em Phys. Rev. Lett.}, 110:181101, May 2013.

\bibitem{PhysRevLett.126.041102}
James Lough, Emil Schreiber, Fabio Bergamin, Hartmut Grote, Moritz Mehmet,
  Henning Vahlbruch, Christoph Affeldt, Marc Brinkmann, Aparna Bisht, Volker
  Kringel, Harald L\"uck, Nikhil Mukund, Severin Nadji, Borja Sorazu, Kenneth
  Strain, Michael Weinert, and Karsten Danzmann.
\newblock First demonstration of 6 db quantum noise reduction in a kilometer
  scale gravitational wave observatory.
\newblock {\em Phys. Rev. Lett.}, 126:041102, Jan 2021.

\bibitem{meers1988recycling}
Brian~J Meers.
\newblock Recycling in laser-interferometric gravitational-wave detectors.
\newblock {\em Physical Review D}, 38(8):2317, 1988.

\bibitem{Morrison:94}
Euan Morrison, Brian~J. Meers, David~I. Robertson, and Henry Ward.
\newblock Automatic alignment of optical interferometers.
\newblock {\em Appl. Opt.}, 33(22):5041--5049, Aug 1994.

\bibitem{Morrison:94_2}
Euan Morrison, Brian~J. Meers, David~I. Robertson, and Henry Ward.
\newblock Experimental demonstration of an automatic alignment system for
  optical interferometers.
\newblock {\em Appl. Opt.}, 33(22):5037--5040, Aug 1994.

\bibitem{PDH}
R.~W.~P. Drever, J.~L. Hall, F.~V. Kowalski, J.~Hough, G.~M. Ford, A.~J.
  Munley, and H.~Ward.
\newblock Laser phase and frequency stabilization using an optical resonator.
\newblock {\em Applied Physics B}, 31(2):97--105, 1983.

\bibitem{Grote:2003ypa}
Hartmut Grote.
\newblock {\em {Making it Work: Second Generation Interferometry in GEO600!}}
\newblock PhD thesis, Hannover U., 2003.

\bibitem{PhysRevD.101.102006}
N.~Mukund, J.~Lough, C.~Affeldt, F.~Bergamin, A.~Bisht, M.~Brinkmann,
  V.~Kringel, H.~L\"uck, S.~Nadji, M.~Weinert, and K.~Danzmann.
\newblock Bilinear noise subtraction at the geo 600 observatory.
\newblock {\em Phys. Rev. D}, 101:102006, May 2020.

\bibitem{bisht2020modulated}
A~Bisht, M~Prijatelj, J~Leong, E~Schreiber, C~Affeldt, M~Brinkmann, S~Doravari,
  H~Grote, V~Kringel, J~Lough, et~al.
\newblock Modulated differential wavefront sensing: alignment scheme for beams
  with large higher order mode content.
\newblock {\em Galaxies}, 8(4):81, 2020.

\bibitem{HochSchm97}
Sepp Hochreiter and J\"urgen Schmidhuber.
\newblock Long short-term memory.
\newblock {\em Neural Computation}, 9(8):1735--1780, 1997.

\bibitem{LSTM_sysID}
Yu~Wang.
\newblock A new concept using lstm neural networks for dynamic system
  identification.
\newblock In {\em 2017 American Control Conference (ACC)}, pages 5324--5329,
  2017.

\bibitem{donahue2015long}
Jeffrey Donahue, Lisa Anne~Hendricks, Sergio Guadarrama, Marcus Rohrbach,
  Subhashini Venugopalan, Kate Saenko, and Trevor Darrell.
\newblock Long-term recurrent convolutional networks for visual recognition and
  description.
\newblock In {\em Proceedings of the IEEE conference on computer vision and
  pattern recognition}, pages 2625--2634, 2015.

\bibitem{CLDNN}
Tara~N. Sainath, Oriol Vinyals, Andrew Senior, and Ha?im Sak.
\newblock Convolutional, long short-term memory, fully connected deep neural
  networks.
\newblock In {\em 2015 IEEE International Conference on Acoustics, Speech and
  Signal Processing (ICASSP)}, pages 4580--4584, 2015.

\bibitem{bozinovski1976influence}
Stevo Bozinovski and Ante Fulgosi.
\newblock The influence of pattern similarity and transfer learning upon
  training of a base perceptron b2.
\newblock In {\em Proceedings of Symposium Informatica}, volume~3, pages
  121--126, 1976.

\bibitem{pratt1991direct}
Lorien~Y Pratt, Jack Mostow, Candace~A Kamm, Ace~A Kamm, et~al.
\newblock Direct transfer of learned information among neural networks.
\newblock In {\em Aaai}, volume~91, pages 584--589, 1991.

\bibitem{PhysRevD.104.064051}
Shreejit Jadhav, Nikhil Mukund, Bhooshan Gadre, Sanjit Mitra, and Sheelu
  Abraham.
\newblock Improving significance of binary black hole mergers in advanced ligo
  data using deep learning: Confirmation of gw151216.
\newblock {\em Phys. Rev. D}, 104:064051, Sep 2021.

\bibitem{iandola2016squeezenet}
Forrest~N Iandola, Song Han, Matthew~W Moskewicz, Khalid Ashraf, William~J
  Dally, and Kurt Keutzer.
\newblock Squeezenet: Alexnet-level accuracy with 50x fewer parameters and< 0.5
  mb model size.
\newblock {\em arXiv preprint arXiv:1602.07360}, 2016.

\bibitem{googlenet}
Christian Szegedy, Wei Liu, Yangqing Jia, Pierre Sermanet, Scott Reed, Dragomir
  Anguelov, Dumitru Erhan, Vincent Vanhoucke, and Andrew Rabinovich.
\newblock Going deeper with convolutions.
\newblock In {\em 2015 IEEE Conference on Computer Vision and Pattern
  Recognition (CVPR)}, pages 1--9, 2015.

\bibitem{szegedy2017inception}
Christian Szegedy, Sergey Ioffe, Vincent Vanhoucke, and Alexander~A Alemi.
\newblock Inception-v4, inception-resnet and the impact of residual connections
  on learning.
\newblock In {\em Thirty-first AAAI conference on artificial intelligence},
  2017.

\bibitem{courbariaux2014training}
Matthieu Courbariaux, Yoshua Bengio, and Jean-Pierre David.
\newblock Training deep neural networks with low precision multiplications.
\newblock {\em arXiv preprint arXiv:1412.7024}, 2014.

\bibitem{han2015deep}
Song Han, Huizi Mao, and William~J Dally.
\newblock Deep compression: Compressing deep neural networks with pruning,
  trained quantization and huffman coding.
\newblock {\em arXiv preprint arXiv:1510.00149}, 2015.

\bibitem{kapurch2010nasa}
Stephen~J Kapurch.
\newblock {\em NASA systems engineering handbook}.
\newblock Diane Publishing, 2010.

\bibitem{van1994n4sid}
Peter Van~Overschee and Bart De~Moor.
\newblock N4sid: Subspace algorithms for the identification of combined
  deterministic-stochastic systems.
\newblock {\em Automatica}, 30(1):75--93, 1994.

\bibitem{boyd1994linear}
Stephen Boyd, Laurent El~Ghaoui, Eric Feron, and Venkataramanan Balakrishnan.
\newblock {\em Linear matrix inequalities in system and control theory}.
\newblock SIAM, 1994.

\bibitem{ljung1998system}
Lennart Ljung.
\newblock System identification.
\newblock In {\em Signal analysis and prediction}, pages 163--173. Springer,
  1998.

\bibitem{franklin2002feedback}
Gene~F Franklin, J~David Powell, Abbas Emami-Naeini, and J~David Powell.
\newblock {\em Feedback control of dynamic systems}, volume~4.
\newblock Prentice hall Upper Saddle River, 2002.

\bibitem{bruinsma1990fast}
N.A. Bruinsma and M.~Steinbuch.
\newblock A fast algorithm to compute the h infinity norm of a transfer
  function matrix.
\newblock {\em Systems \& Control Letters}, 14(4):287--293, 1990.

\bibitem{Apkarian2006}
P.~Apkarian and D.~Noll.
\newblock Nonsmooth h-infinty synthesis.
\newblock {\em IEEE Transactions on Automatic Control}, 51(1):71--86, 2006.

\bibitem{aastrom2006advanced}
Karl~Johan {\AA}str{\"o}m, Tore H{\"a}gglund, and Karl~J Astrom.
\newblock {\em Advanced PID control}, volume 461.
\newblock ISA-The Instrumentation, Systems, and Automation Society Research
  Triangle Park, 2006.

\bibitem{sutton2018reinforcement}
Richard~S Sutton and Andrew~G Barto.
\newblock {\em Reinforcement learning: An introduction}.
\newblock MIT press, 2018.

\bibitem{haarnoja2018soft}
Tuomas Haarnoja, Aurick Zhou, Pieter Abbeel, and Sergey Levine.
\newblock Soft actor-critic: Off-policy maximum entropy deep reinforcement
  learning with a stochastic actor.
\newblock In {\em International conference on machine learning}, pages
  1861--1870. PMLR, 2018.

\bibitem{sutton1999policy}
Richard~S Sutton, David McAllester, Satinder Singh, and Yishay Mansour.
\newblock Policy gradient methods for reinforcement learning with function
  approximation.
\newblock {\em Advances in neural information processing systems}, 12, 1999.

\bibitem{mnih2016asynchronous}
Volodymyr Mnih, Adria~Puigdomenech Badia, Mehdi Mirza, Alex Graves, Timothy
  Lillicrap, Tim Harley, David Silver, and Koray Kavukcuoglu.
\newblock Asynchronous methods for deep reinforcement learning.
\newblock In {\em International conference on machine learning}, pages
  1928--1937. PMLR, 2016.

\bibitem{schulman2015high}
John Schulman, Philipp Moritz, Sergey Levine, Michael Jordan, and Pieter
  Abbeel.
\newblock High-dimensional continuous control using generalized advantage
  estimation.
\newblock {\em arXiv preprint arXiv:1506.02438}, 2015.

\bibitem{gu2016q}
Shixiang Gu, Timothy Lillicrap, Zoubin Ghahramani, Richard~E Turner, and Sergey
  Levine.
\newblock Q-prop: Sample-efficient policy gradient with an off-policy critic.
\newblock {\em arXiv preprint arXiv:1611.02247}, 2016.

\bibitem{ota2021training}
Kei Ota, Devesh~K Jha, and Asako Kanezaki.
\newblock Training larger networks for deep reinforcement learning.
\newblock {\em arXiv preprint arXiv:2102.07920}, 2021.

\bibitem{wiering2008ensemble}
Marco~A Wiering and Hado Van~Hasselt.
\newblock Ensemble algorithms in reinforcement learning.
\newblock {\em IEEE Transactions on Systems, Man, and Cybernetics, Part B
  (Cybernetics)}, 38(4):930--936, 2008.

\bibitem{vaswani2017attention}
Ashish Vaswani, Noam Shazeer, Niki Parmar, Jakob Uszkoreit, Llion Jones,
  Aidan~N Gomez, {\L}ukasz Kaiser, and Illia Polosukhin.
\newblock Attention is all you need.
\newblock {\em Advances in neural information processing systems}, 30, 2017.

\bibitem{reed2022generalist}
Scott Reed, Konrad Zolna, Emilio Parisotto, Sergio~Gomez Colmenarejo, Alexander
  Novikov, Gabriel Barth-Maron, Mai Gimenez, Yury Sulsky, Jackie Kay,
  Jost~Tobias Springenberg, et~al.
\newblock A generalist agent.
\newblock {\em arXiv preprint arXiv:2205.06175}, 2022.

\bibitem{brohan2022rt}
Anthony Brohan, Noah Brown, Justice Carbajal, Yevgen Chebotar, Joseph Dabis,
  Chelsea Finn, Keerthana Gopalakrishnan, Karol Hausman, Alex Herzog, Jasmine
  Hsu, et~al.
\newblock Rt-1: Robotics transformer for real-world control at scale.
\newblock {\em arXiv preprint arXiv:2212.06817}, 2022.

\end{thebibliography}

\end{document}